\documentclass[preprint,12pt]{elsarticle}%
\usepackage{epsfig}
\usepackage{amsmath}
\usepackage{graphicx,color}
\usepackage{amsmath}
\usepackage{amsfonts}
\usepackage{amssymb}
\usepackage{graphicx}%
\graphicspath{{./img/}}

\begin{document}
\begin{frontmatter}
\title{Fluctuations and the rapidity dependence \\
of charged particles spectra \\
in fixed centrality bins in p$A$ collisions}
\author[bnl,ccnu]{Larry McLerran}
\author[jag]{Michal Praszalowicz}
\address[bnl]{Physics Dept, Bdg. 510A, Brookhaven National Laboratory, Upton, NY-11973, USA}
\address[ccnu]{Physics Dept, China Central Normal University, Wuhan, China}
\address[jag]{M. Smoluchowski Institute of Physics, Jagiellonian University, \\
ul. S. {\L}ojasiewicza 11, 30-348 Krak{\'o}w, Poland.}
\vspace{0.3cm}
\begin{abstract}
We argue that large fluctuations in the saturation momentum are necessary to explain the ATLAS
and ALICE
data on p$A$ collisions measured at the LHC.  Using a form for the distribution of
fluctuations motivated by theoretical studies
of the non-linear evolution equations for the Color Glass Condensate, we find a remarkably good agreement
between theory and the measured distributions.  If the saturation momentum fluctuates,
we argue that the cross section for a proton probe should also fluctuate, consistent with previous
observations.
\end{abstract}
\end{frontmatter}

\section{Introduction}

\begin{figure}[h]
\centering
\includegraphics[height=4.0cm]{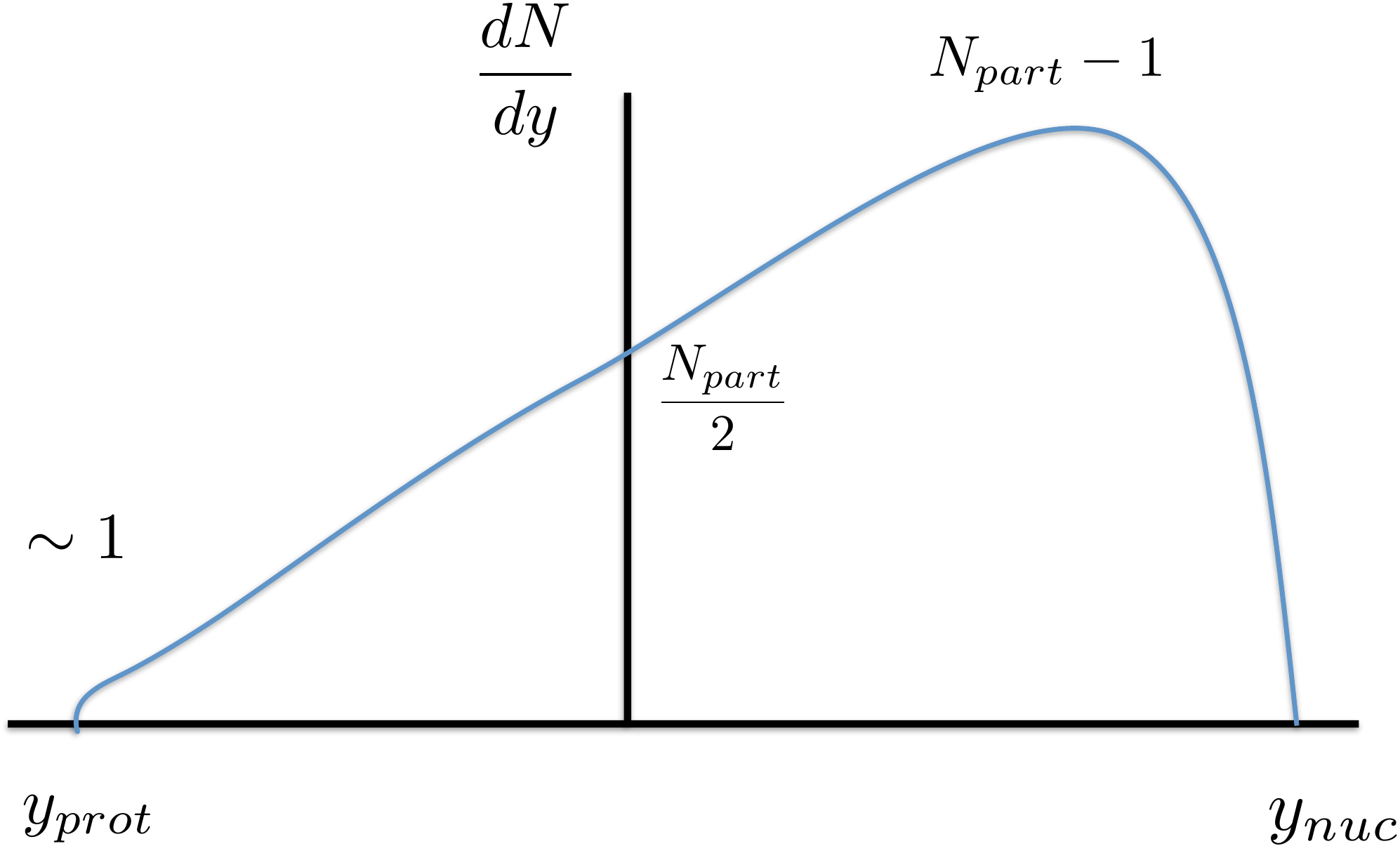}\caption{An illustration of the
multiplicity distribution of produced particles in p$A$ collisions as a
function of rapidity in the simplest version of the wounded nucleon model.}%
\label{fig:spectraWN}%
\end{figure}

In a recent paper by Bzdak and Skokov~\cite{Bzdak:2013zla}, it was argued that
the number of participants dependence of the multiplicity as a function of
centrality provided a crucial test which might discriminate between the
description of heavy ion collisions provided by the Color Glass
Condensate\cite{MLV} and that of the wounded nucleon
model\cite{Bialas:1976ed,Bialas:2004su}. Their basic idea is very simple: In
the simplest version of the wounded nucleon model, the rapidity dependence of
the multiplicity distribution for a collision of a projectile nucleon against
$N_{\mathrm{part}}-1$ nucleons in the target is a simple triangular
distribution going from a number proportional the number or participants for
rapidities close to that of the nucleus to a number of order 1 at rapidities
corresponding to the beam nucleon. This is shown in Fig.~\ref{fig:spectraWN}

\begin{figure}[h]
\centering
\includegraphics[height=4.0cm]{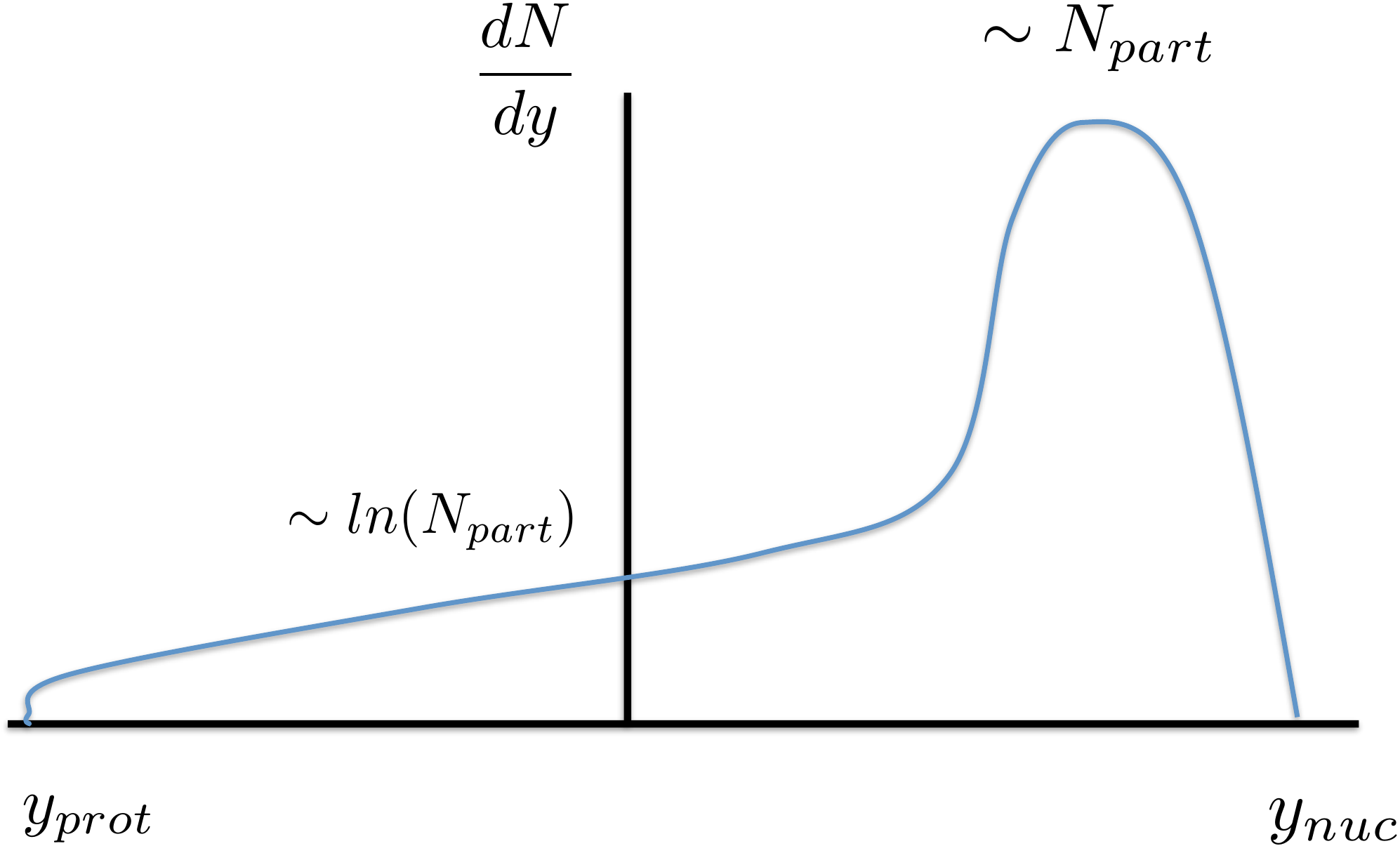}\caption{An illustration of the
multiplicity distribution of produced particles in p$A$ collisions as a
function of rapidity in the simplest version of a model for the Color Glass
Condensate.}%
\label{fig:spectraCG}%
\end{figure}

On the other hand, the simplest version of the Color Glass Condensate model is
that the rapidity distribution in the fragmentation of the target nucleus is
of order $N_{\mathrm{part}}$ but elsewhere of order 1 up to logarithms of the
number of participants. This is because for a probe which has a Lorentz gamma factor 
of order  $\gamma \ge A^{1/3}$ relative to the nucleus, the nucleons in the
nucleus act coherently, and for the typical transverse momentum scale
associated with particle production, the nucleus appears as a black disk. The
multiplicity is controlled by the number of partons in the proton evolved to
the saturation scale of the nucleus. In fact there is a mild logarithmic
correction to this, so the multiplicity scales as $\ln N_{\mathrm{part}}$
\cite{Kovchegov:1998bi,Dumitru:2001ux}. Schematically the Color Glass
Condensate description is shown in Fig.~\ref{fig:spectraCG}. 

The number of participants dependence of the pseudorapidity distributions in
p$A$ collisions at the LHC energy has been recently measured by the ATLAS
\cite{atlas} and ALICE \cite{Adam:2014qja} collaborations. In ATLAS, they
extract the number of participant dependence of the integrated multiplicity
under a variety of different assumptions concerning the cross-section for the
proton penetrating through the nucleus. These correspond to a fixed Glauber
cross section for the proton and to two different parameterizations of the
Glauber-Gribov description \cite{Guzey:2005tk,Alvioli:2013vk}. The
Glauber-Gribov model allows the proton cross-section to vary by Gausssian
fluctuations. Similar procedures have been also applied by the ALICE experiment.

In the analysis of the next section, we compare expectations from the naive
versions of the wounded nucleon model and that of the Color Glass Condensate
to the experimental data. This analysis consists of comparing the
$N_{\mathrm{part}}$ dependence predicted by the two models as a function of
rapidity to that of the experimental data. Neither model gives an acceptable
description of the data for any extraction of the number of participant dependence.

How might one resolve this impasse? A hint is suggested by the extraction of
the relationship between number of participants and multiplicity used to
determine centrality. One might argue that in the fragmentation region of the
nucleus, all models agree that the multiplicity dependence is linear in the
number of participants. If so, the closest we come to the fragmentation region
with the data is for $2 < \eta< 2.7$. While this is not forward enough to
really be in the nuclear fragmentation region, we see that the data favors
large fluctuations in the proton cross section. This suggests there might be
fluctuations in the proton saturation momentum.

 Before we discuss fluctuations of the saturation scale,
let us remark that there exists a number of theoretical papers that took
an effort to describe the p$A$ data within the framework of the different
versions of  the CGC-based models. A complete list of such models can
be found in a review paper \cite{Albacete:2013ei} that was, however, 
published before the first LHC run. More recently successful
description of the p$A$ data has been presented in 
Refs.~\cite{Rezaeian:2012ye,Schenke:2015aqa,Dusling:2014oha,Schenke:2013dpa}
and references therein. 

One expects a relationship between the cross section and the
saturation momentum. If we require that the local density of gluons, denoted
as $\Lambda^{2}$, in an impact parameter model of the proton factorizes as
proportional to the number of gluons per unit area on average times an impact
parameter profile, which for simplicity can be assumed to be exponential, then%

\begin{equation}
\Lambda^{2} = Q_{\mathrm{s}}^{2} e^{-\mu R}%
\end{equation}
Requiring that the cross-section corresponds to the radius at which this
density has some fixed value $\Lambda_{0}^{2}$, perhaps $\Lambda
^{2}_{\mathrm{QCD}} $, we conclude that the radius of the proton grows as
\begin{equation}
R \sim{\frac{1 }{\mu}} \ln(Q_{\mathrm{s}}^{2}/\Lambda_{0}^{2})
\end{equation}
Therefore the cross section $\sigma\propto\ln^{2}(Q_{\mathrm{s}}^{2}%
/\Lambda^{2})$ will fluctuate if the saturation momentum does. There are of
course many weaknesses of this analysis, perhaps the most important one is
that we do not really know the impact parameter distribution for matter inside
the proton, so that the exponential form, while valid at very large radii, may
be misleading.

The emergence of the saturation scale and
geometrical scaling in the theory of the Color Glass Condensate has been
introduced in Refs.~\cite{Munier:2003sj,Munier:2003vc}. 
It is based on the
fact that the nonlinear evolution equations in small Bjorken $x$ regime
\cite{jimwlk,BK} possess traveling wave solutions \cite{Munier:2003sj,Munier:2003vc}.
It has been argued, however, in Ref.~\cite{Iancu:2004es} and expanded in Ref.~\cite{Hatta:2006hs}
that the scaling CGC solutions have exponential tails which are unphysical, and therefore have 
to be modified. This modification is based on the analogy with the statistical physics, and has clear
interpretation in the dipole model, where the small tail of the amplitude is essentially counting
a number of interacting dipoles. This discrete nature of the tail introduces stochastic diffusion
that affects the shape of the entire amplitude, which -- event by event --
satisfies geometrical scaling, with the
saturation scale that is, however, somewhat different than the one obtained in the mean field approximation.
It has been
argued that these fluctuations are Gaussian in the logarithm of the saturation momentum
\cite{Iancu:2004es}, and controlled by a probability distribution:
\begin{align}
P(\rho)  & =\frac{1}{\sqrt{2\pi}\sigma}\exp\left(  -\frac{(\ln Q_{\mathrm{s}}^{2}%
/Q_{0}^{2}-\ln Q_{{p}}^{2}/Q_{0}^{2})^{2}}{2\sigma^{2}}\right) \nonumber\\
& =\frac{1}{\sqrt{2\pi}\sigma}\exp\left(  -\frac{\rho^{2}}{2\sigma^{2}}\right)
\label{distr}%
\end{align}
where%
\begin{equation}
\rho=\ln\frac{Q_{\mathrm{s}}^{2}}{Q_{{p}}^{2}}.\label{rho}%
\end{equation}
Here $Q_{\mathrm{s}}^{2}$ is the proton saturation momentum fluctuating around its
  logarithmic average denoted as $\ln Q_{{p}}^{2}$ (with $Q_0^2 \sim 1$~GeV$^2/c^2$ 
  being an arbitrary momentum scale, which cancels out in (\ref{distr})). 
  
 It has been furthermore argued in \cite{Iancu:2004es} that the width $\sigma$ 
of the distribution (\ref{distr}) is growing with the energy of the collision and
depends upon the rapidity where measurements are made. One has to stress,
however, that this result is only asymptotic, and -- as the authors of Ref.~\cite{Iancu:2004es}
admit explicitly -- there is no general analytical proof of this result and its status is still that of
a conjecture. Phenomenologically scaling violations corresponding ro
\begin{equation}
\sigma\sim\sqrt{y}\;\; \text{with}\;\; y=\log(1/x).
\label{sigma_y}
\end{equation}
have been searched for in the DIS data with negative result
\cite{Gelis:2006bs,Beuf:2008mf,Praszalowicz:2013iyi}.

There are two possible ways out from this situation. Either
(\ref{sigma_y}) holds down to the present energies (or presently accessible
Bjorken $x$'s), in which case the distribution (\ref{distr}) collapses to
$\delta(\ln Q_{\mathrm{s}}^{2}%
/ Q_{{p}}^{2})$, or result
(\ref{sigma_y}) is only asymptotic, and for low energies $\sigma$
tends to a constant. As we shall show in the following, in such a case 
geometrical scaling would still hold.

To this end let us consider a DIS amplitude $T$ that
exhibits geometrical scaling with respect to the fluctuating saturation
momentum $Q_{\mathrm{s}}(x)$:%
\begin{equation}
T(Q^{2}/Q_{s}^{2})=T(\rho_{Q}-\rho)
\end{equation}
where $\rho_{Q}=\ln(Q^{2}/Q_{0}^{2})$. Note that $T$ depends on $y=\ln(1/x) $
through $x$-dependence of $Q_{\mathrm{s}}$. For illustrative purposes let's assume,
following Ref.~\cite{Hatta:2006hs}, that%
\begin{equation}
T(\rho_{Q}-\rho)=\Theta(\rho-\rho_{Q})\label{TTheta}%
\end{equation}
representing a step-like front moving with rapdity $y=\ln(1/x)$, which plays a
role of an evolution time. Averaging (\ref{TTheta}) with probability
distribution (\ref{distr}) gives:%
\begin{align}
\left\langle T_{x}(\rho_{Q})\right\rangle  & =%
{\displaystyle\int\limits_{-\infty}^{\infty}}
d\rho\,P(\rho)\,T(\rho_{Q}-\rho))\,\nonumber\\
& =\frac{1}{2}\operatorname*{Erfc}\left(  \frac{\log(Q^{2}/Q_{{p}}^{2}(x))}%
{\sqrt{2}\sigma}\right)  .
\end{align}
So the new scaling variable is%
\begin{equation}
\log\tau=\frac{1}{\sigma}\log(Q^{2}/Q_{{p}}^{2}(x)).
\end{equation}
It is also straightforward to check that if $T$ 
is a function of $\rho_Q-\rho$ then, after integrating over fluctuations,
$T \rightarrow \left< T(\rho_{p}-\rho_Q) \right> $ so that geometrical 
scaling is preserved 
so long as the width of fluctuations is rapidity independent.

We do not have any explicit model leading to the fluctuations with
the constant width, although -- given the phenomenological arguments above -- such fluctuations
are compatible with the DIS data.  The fluctuations of the saturation momentum 
we are employing here are caused by
an evolution with energy of the initial conditions of the forward scattering
amplitude and are associated with the low density region (\emph{i.e.} with a
region where the amplitude is small, which means rather large Bjorken $x$'s). But for
large $x$'s parton distributions are non-perturbative, and the fluctuations
can be of order one. In contrast, it was established that pomeron loops
effects  \cite{Dumitru:2007ew}, once at small $x$, are small, and therefore they cause only a small
change in the scale of rapidity fluctuations as one goes to higher and higher
energies. Large
nonperturbative fluctuations in turn, propagate to small $x$'s \emph{i.e.} to
the region we are interested in.

In the following  shall ignore the
rapidity dependence and treat $\sigma$ as a constant at fixed LHC energy.
Obviously one needs to include beam energy dependence if this result would be
used over a very wide range of energy from RHIC to the LHC. 

In the analysis of p$A$ collisions which we present below, we will consider
fluctuations only of the nucleon. 
Including the fluctuations of the saturation scale in the nucleus might be done, but we  expect
a small effect.  This is because the fractional change in the saturation momentum squared 
of the nucleus should be reduced by
random walks.  We might expect $\delta Q^2_\mathrm{s}/Q_\mathrm{s}^2 \sim 1/\sqrt{N_{A\,{\rm part}}}$ where 
$N_{A\,{\rm part}}$ is the number of participating nucleons in the
target nucleus.
 In pp collisions, we would need 
to consider fluctuations in both
protons. Without such inclusion of the effects of fluctuations of the nuclear
saturation momentum, the results we present here should be taken to
demonstrate qualitatively and semi-quantitatively the effects of fluctuating
saturation momenta, and quantitative conclusions should await a more
systematic analysis.

\section{Limitations of the Wounded Nucleon and Color Glass Descriptions of
the LHC pPb Data}

\begin{figure}[h]
\centering
\includegraphics[height=7.5cm]{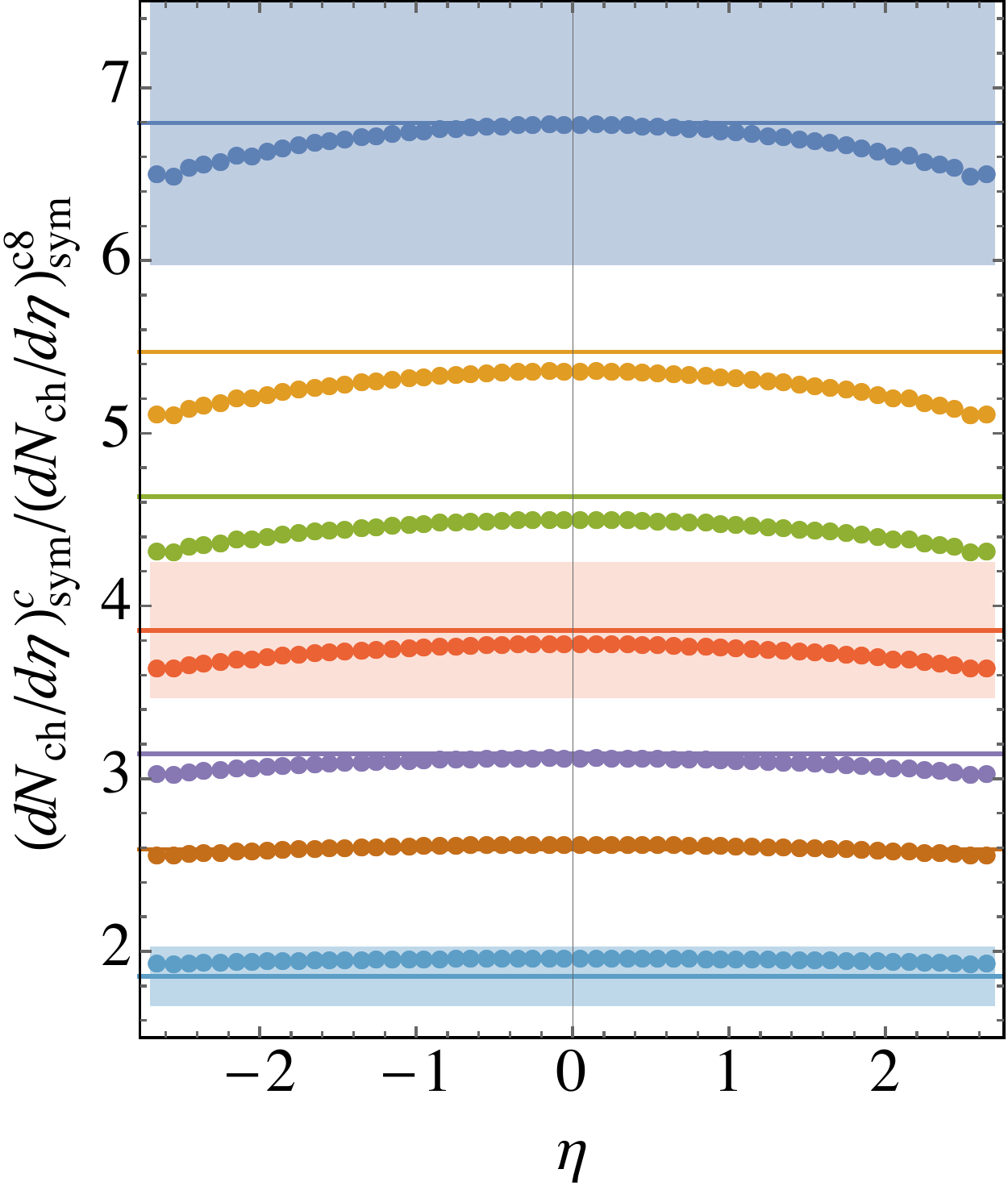}~~~
\includegraphics[height=7.6cm]{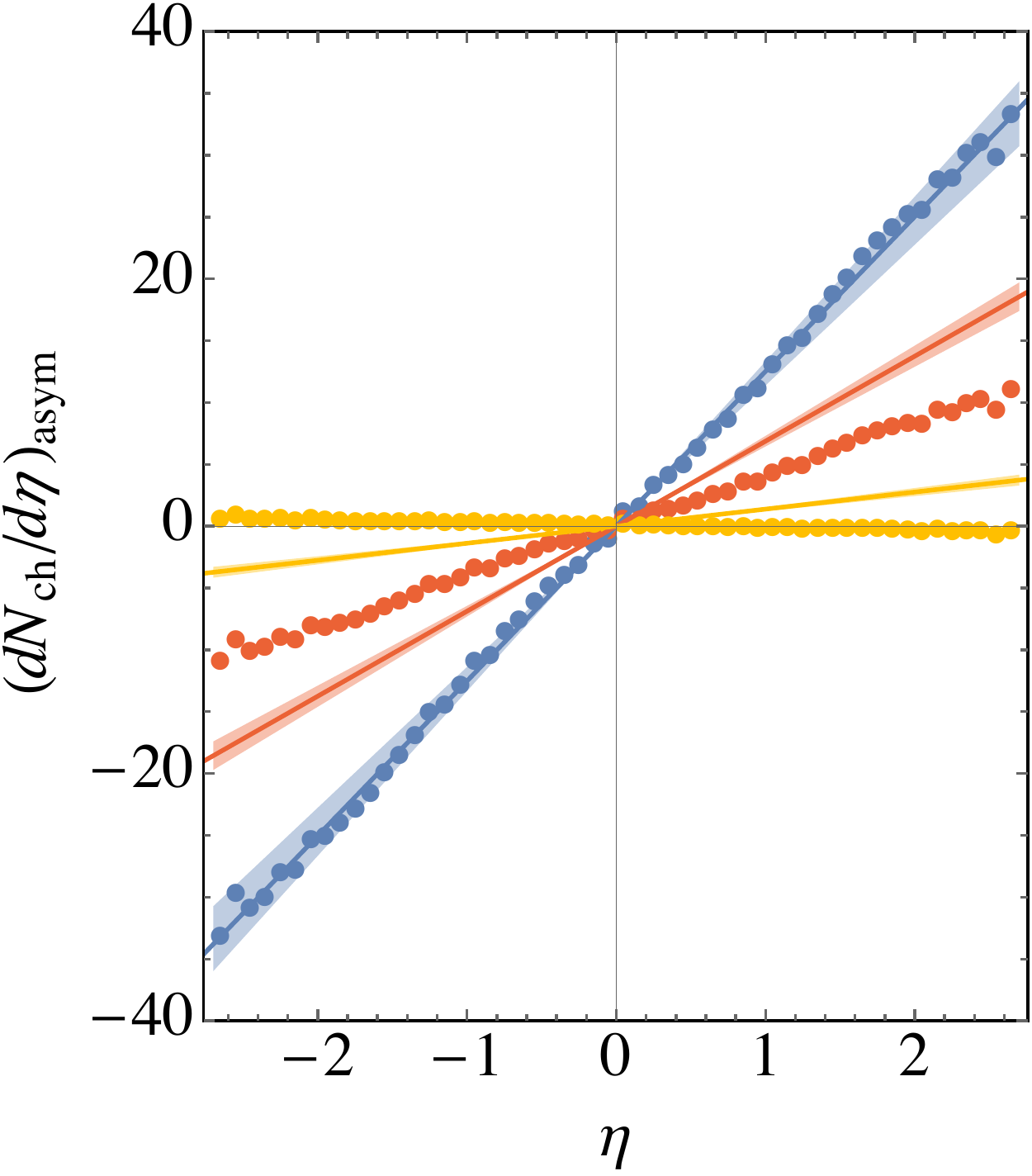}\caption{Wounded nucleon
model predictions for symmetrized (left) and antisymmetrized (right) spectra
defined in Eqs.~(\ref{sym}) and (\ref{asym}). In the left panel ratios of
symmetrized distributions to the 60--90~\% (c8) centrality class are plotted
together with the straight lines corresponding to the ratios of respective
$N_{\mathrm{part}}$. Error bands due to the uncertainty of $N_{\mathrm{part}}$
are for clarity plotted only for three centrality classes: 0--1~\% (c1 -- top,
blue), 10--20~\% (c4 -- middle, orange) and 40--50~\% (c7 -- bottom, light
blue). In the right panel antisymmetrized spectra (\ref{asym}) are plotted for
three centrality classes 0--1~\% (blue), 10--20~\% (orange) and 60--90~\%
(yellow). Shaded bands correspond to the uncertainty of $N_{\mathrm{part}}$.
Experimental points are from ATLAS Collaboration.}%
\label{SymAsym}%
\end{figure}

In the wounded nucleon (WN) model \cite{Bialas:1976ed,Bialas:2004su}
multiplicity is given in terms of the number of wounded nucleons multiplied by
a function describing radiation of particles by one struck participant:%
\begin{equation}
\frac{dN}{d\eta}=w_{p}F_{p}(\eta)+w_{A}F_{A}(\eta).
\end{equation}
Total number of participants is therefore $N_{\mathrm{part}}=w_{p}%
+w_{A}=1+w_{A}$. The only difference between functions $F_{p}$ and $F_{A}$ is
that they radiate particles into different hemispheres. Therefore%
\begin{equation}
F_{p}(\eta)=F(\eta),\;F_{A}(\eta)=F(-\eta).\;
\end{equation}
In order to test WN model against data it is convenient to introduce
symmetrized and antisymmetrized spectra \cite{Bialas:2004su}:%
\begin{align}
\left.  \frac{dN}{d\eta}\right\vert _{\text{sym}}  & =\frac{dN}{dy}%
(\eta)+\frac{dN}{dy}(-\eta)=(w_{p}+w_{A})\left(  F(\eta)+F(-\eta)\right)
,\label{sym}\\
\left.  \frac{dN}{d\eta}\right\vert _{\text{asym}}  & =\frac{dN}{d\eta}%
(\eta)-\frac{dN}{d\eta}(-\eta)=(w_{p}-w_{A})\left(  F(\eta)-F(-\eta)\right)
.\label{asym}%
\end{align}
Equation (\ref{sym}) implies that ratios of symmetrized spectra at different
centralities are independent of $\eta$ and are equal to the ratio of number of
participants. Ratios of symmetrized spectra to the centrality class with
lowest number of participants are plotted in the left panel of
Fig.~\ref{SymAsym}. For clarity only in three cases error bands due to the
uncertainties of $N_{\mathrm{part}}$ are also displayed. We can see that an
overall magnitude of these ratios agrees with the data, however -- especially
for large multiplicities -- they are not $\eta$ independent in contrast to the
prediction of the WN model.

On the other hand antisymmetrized spectra are proportional to $N_{\text{part}%
}-1$, assuming that on the proton side there was only one participant. We have
fit the slope of the antisymmetrized spectra (which are to a good accuracy
straight lines) to the highest $N_{\mathrm{part}}$ data, obtaining predictions
for other centralities. This is plotted in the right panel of
Fig.~\ref{SymAsym} for three centrality classes. Shaded bands around the
theoretical lines correspond again to the uncertainties of $N_{\mathrm{part}}
$. We see that here the wounded nucleon model fails to describe the data and
that slopes of the antisymmetrized spectra have some additional dependence on
$N_{\mathrm{part}}$.

In the case of CGC, multiplicity is given in terms of the saturation
scales\cite{Kharzeev:2004if}:
\begin{equation}
\frac{dN}{dy} = S_{\bot}Q_{p}^{2}\left(  2+\ln\frac{Q_{A}^{2}}{Q_{p}^{2}%
}\right) \label{mult1}%
\end{equation}
and the multiplicity in p$A$ collisions is totally driven by the proton
saturation scale. Here $S_{\bot}$ is a
transverse area corresponding to an overlap of colliding hadrons. In the case
of heavy ion and p$A$ collisions it has rather well defined geometrical
meaning, whereas in pp (small system) scattering it is a parameter related to
the multiplicity of a given event. As such it may also fluctuate, however in
the present paper such fluctuations are neglected. For fixed impact parameter%
\begin{equation}
\rho_{A}=\ln\frac{Q_{A}^{2}}{Q_{p}^{2}} \sim\ln N_{\text{part}}%
\label{rhoApart1}%
\end{equation}
has only logarithmic dependence on $N_{\text{part}}$.

Formulae \thinspace(\ref{mult1}) and (\ref{rhoApart1}) predict both energy
(denoted hereafter as $W=\sqrt{s}$)
dependence and rapidity (denoted below as $y$) dependence and also $N_{\text{part}}$ dependence of
multiplicities through the dependence of the saturation scales on these
quantities \cite{Kharzeev:2004if}:%
\begin{align}
Q_{p}^{2}(W,y) &  =  Q_{0}^{2}\left(  \frac{W}{W_{0}}\right)  ^{\lambda}%
\exp(\lambda y),\nonumber\\
Q_{A}^{2}(W,y) &  =  Q_{0}^{2}N_{\text{part}}\left(  \frac{W}{W_{0}}\right)
^{\lambda}\exp(-\lambda y)\label{Qydep}%
\end{align}
where we take for $\lambda=0.32$ in agreement with recent analysis
\cite{Praszalowicz:2012zh} of the combined DIS data from HERA. Here $W_0$
is an arbitrary energy scale taking care of the proper dimension of the saturation
scales.

Pseudo-rapidity $\eta$, which is measured, is related to rapidity $y$ in the
following way
\begin{equation}
y(\eta)=\frac{1}{2}\ln\left[  \frac{\sqrt{\eta_{0}^{2}+\sinh^{2}\eta}%
+\sinh\eta}{\sqrt{\eta_{0}^{2}+\sinh^{2}\eta}-\sinh\eta}\right]  \ .
\end{equation}
The corresponding Jacobian reads:
\begin{equation}
h(\eta)=\frac{\cosh\eta}{\sqrt{\eta_{0}^{2}+\sinh^{2}\eta}}%
\end{equation}
where%
\begin{equation}
\eta_{0}^{2}=\frac{m^{2}+p_{\text{T}}^{2}}{p_{\text{T}}^{2}}.
\end{equation}
In the following we shall use $\eta^{2}_{0}=1.35$.

In Fig.~\ref{dNdydatath} we plot theoretical predictions of Eq.~(\ref{mult1})
and the corresponding data from ATLAS and ALICE. One can see that CGC formula
fails to describe overall normalization and the slope (note that theoretical
curves have been normalized at $\eta=0$). In the remainder of this paper we
shall show that inclusion of the fluctuations of the proton saturation scale
is going to improve upon both issues

\section{Fluctuations of the saturation scale}

In this Section we shall derive formulae for the multiplicity of gluons in the
fluctuating case with probability distribution given by Eq.~(\ref{distr}). For
fluctuating saturation scale $Q^{2}<Q^{2}_{A}$ multiplicity takes the form
given by Eq.~(\ref{mult1}) with $Q_{p}^{2}$ replaced by $Q^{2}$, and for
$Q_{A}^{2} < Q^{2}$ we simply interchange $Q_{A} \leftrightarrow Q$. With
this, we have ignored a possible slow dependence upon the saturation momentum
of the proton that affects the transverse area. This is slowly varying, like a
logarithm squared, according to the result in the introduction, but we will
later need to include such an effect to properly normalize the pseudorapidity
distribution that we compute.

Now we can compute the multiplicity as
\begin{align}
{\frac{{dN} }{{dy}}}  & =  S_{\bot} Q_{p}^{2} \left\{  \int_{-\infty}%
^{\rho_{A}} d\rho~ (\rho_{A}-\rho+2)e^{\rho}P(\rho) + \int_{\rho_{A}}^{\infty}
d\rho~ (\rho-\rho_{A} +2) e^{\rho_{A}} P(\rho) \right\} \nonumber\\
&
\end{align}
The first integral reads
\begin{align}
I_{1}  & = \int_{-\infty}^{\rho_{A}} d\rho~ (\rho_{A}-\rho+2) e^{\rho}
P(\rho)\label{i1}\\
&  =  \frac{e^{\sigma^{2}/2}}{\sqrt{2\pi}} \left\{  \sigma e^{-(\rho
_{A}-\sigma^{2})^{2}/2\sigma^{2}} + \sqrt{{\frac{\pi}{2}}} (\rho_{A} -
\sigma^{2} +2) (1 + \operatorname*{Erf}\left( {\frac{{\rho_{A}-\sigma^{2}}
}{{\sqrt{2}\sigma}}}\right)  \right\} \nonumber
\end{align}
and the remaining integral is
\begin{align}
I_{2}  & = \int_{\rho_{A}}^{\infty} d\rho~ (\rho-\rho_{A} +2) e^{\rho_{A}}
P(\rho)\label{i2}\\
&  =  \frac{e^{\rho_{A}}}{\sqrt{2\pi}} \left\{  \sigma e^{-\rho_{A}%
^{2}/2\sigma^{2}} + \sqrt{{\frac{\pi}{2}}} (2-\rho_{A}) (1 -
\operatorname*{Erf}\left( \frac{\rho_{A}}{{\sqrt{2}\sigma} }\right)  \right\}
.\nonumber
\end{align}

This gives
\begin{equation}
{\frac{{dN} }{{dy}}}= S_{\bot} Q_{p}^{2} (I_{1}+I_{2})\label{sumI}%
\end{equation}
The saturation scales entering
Eqs.~(\ref{i1})--(\ref{sumI}) are given by Eq.~(\ref{Qydep}); they depend on
energy and rapidity, and -- in the case of the nucleus -- also on
$N_{\mathrm{part}} $. Note that in the limit where $\sigma\rightarrow0$,
\emph{i.e.} in the limit with no fluctuations, $I_{2} \rightarrow0$ and $I_{1}
\rightarrow(2+\rho_{A})$ in agreement with Eq.~(\ref{mult1}).

\begin{figure}[h]
\centering
\includegraphics[width=6.5cm]{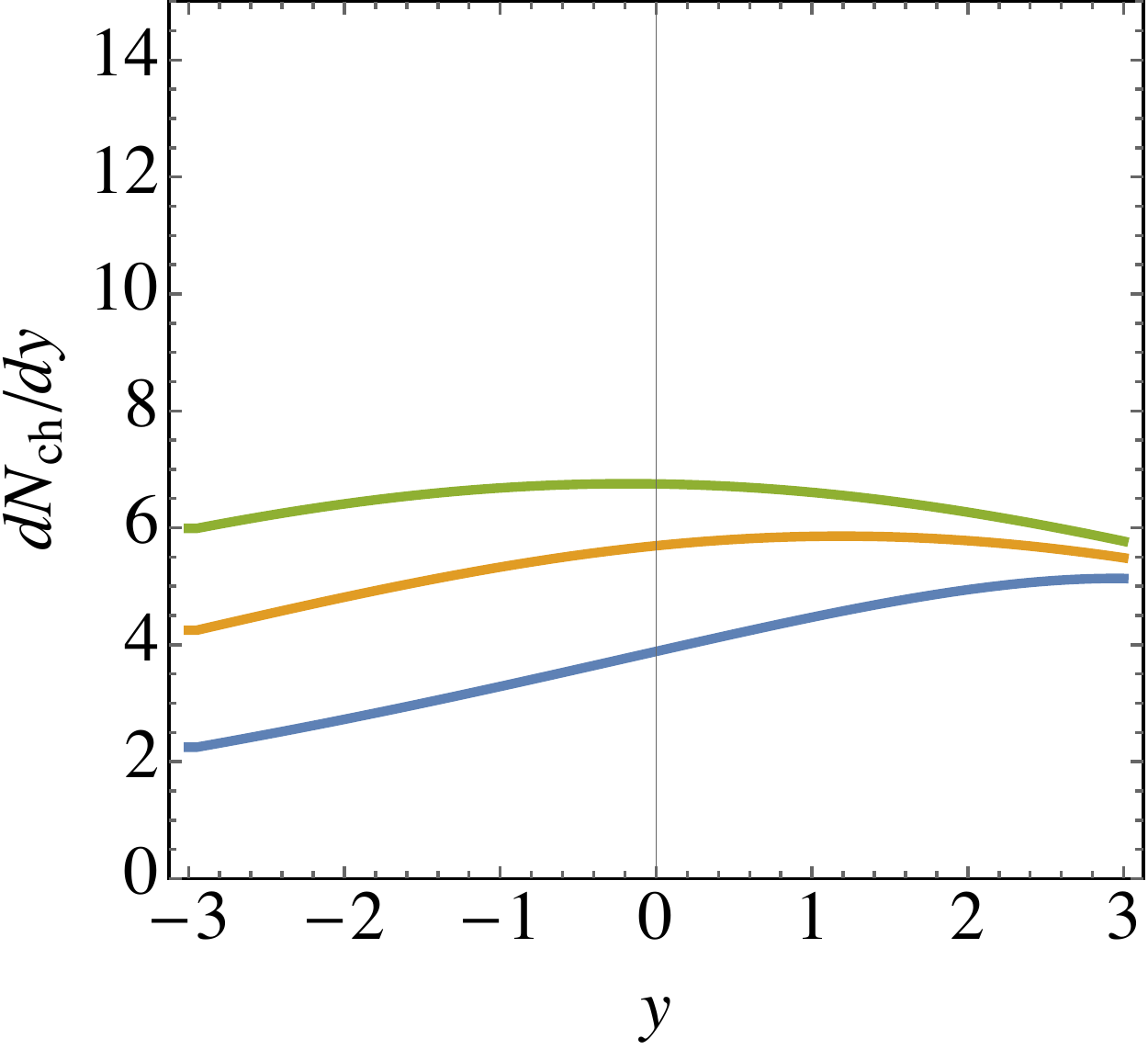}~~~
\includegraphics[width=6.5cm]{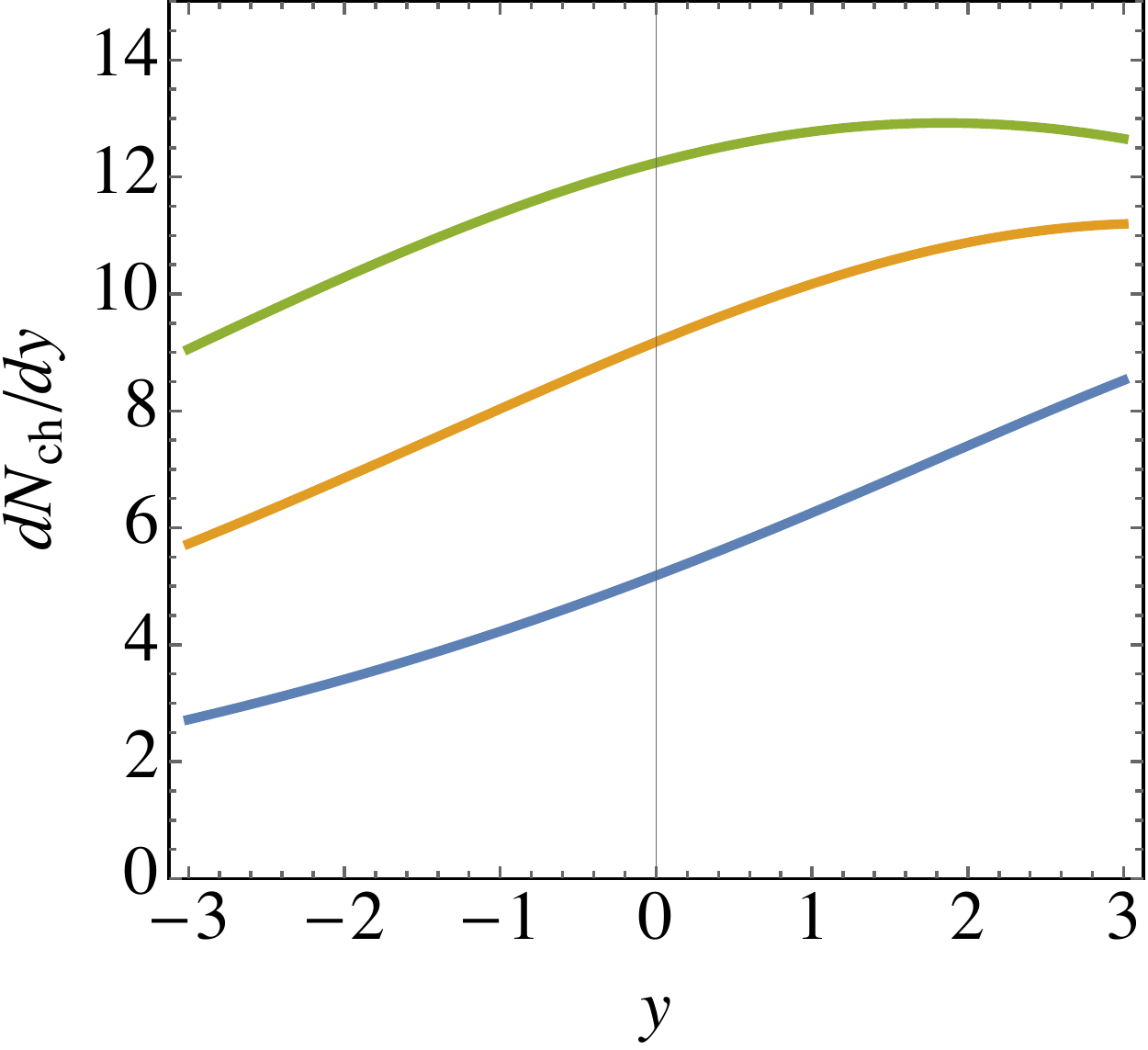}\caption{Multiplicity
dependence on rapidity $y$ for the non-fluctating case (blue) and for the case
with fluctuations with $\sigma=1$ (orange) and 2 (green). Left plot
corresponds to $N_{\mathrm{part}}=6.59$ and the right one to $N_{\mathrm{part}%
}=24.13$. }%
\label{dNdynodata}%
\end{figure}

In order to illustrate the effect of fluctuations we plot in
Fig.~\ref{dNdynodata} the p$A$ multiplicity for two different centrality
classes for the non-fluctuating case (lower blue curve) and for the case with
fluctuations included for $\sigma=1$ (middle orange line) and $\sigma=2$
(upper green line). We see clearly two effects whose strength depends on
centrality: first, an overall normalization is increasing with increasing
$\sigma$ and second, distributions are getting flatter when $\sigma$ is
rising. For small number of participants we observe even a counterintuitive
behavior that multiplicity is smaller on the nucleus side (upper green line in
the left panel of Fig.~\ref{dNdynodata}).

\section{Multiplicity distributions in p$A$}

\begin{figure}[ptbh]
\centering
\includegraphics[width=6.5cm]{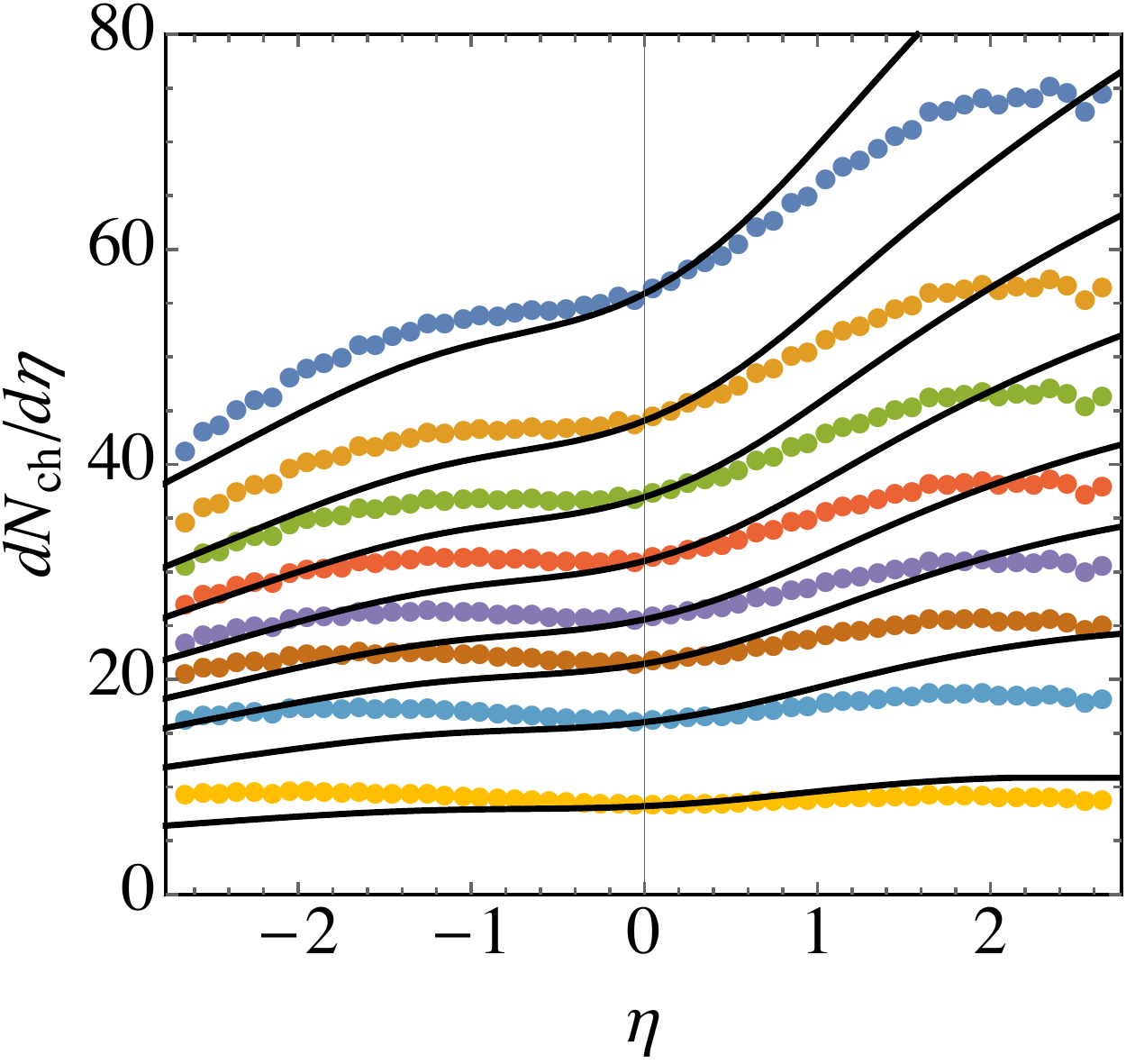}~~~~
\includegraphics[width=6.5cm]{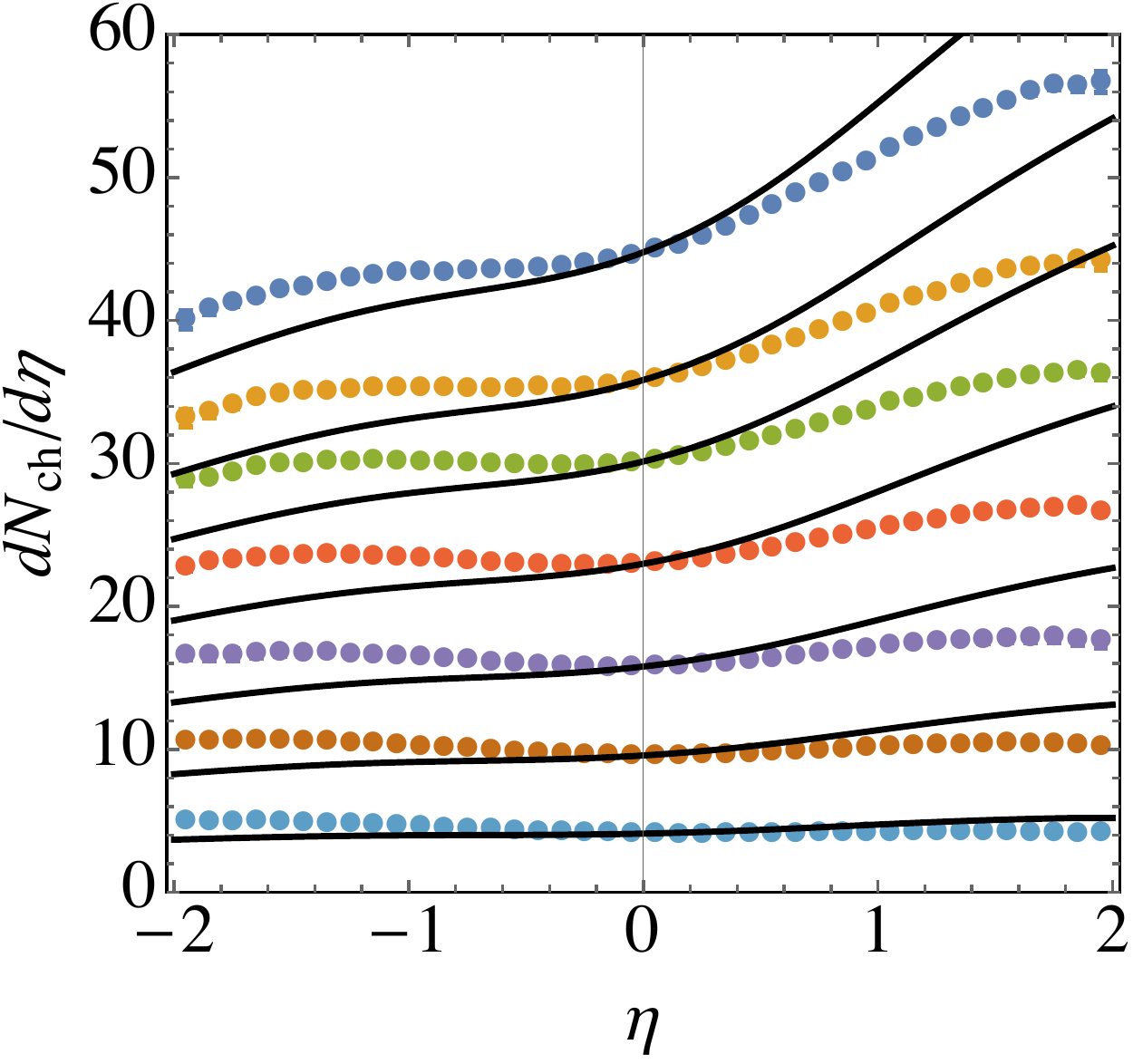}\caption{Multiplicity
dependence on pseudorapidity $\eta$ given by formula (\ref{mult1}) for ATLAS
(left) and ALICE (right) for different centrality classes described in the
text. Theoretical curves have been normalized at $\eta= 0$ (see
Fig.~\ref{NormNpart}).}%
\label{dNdydatath}%
\end{figure}

\begin{figure}[h]
\centering
\includegraphics[width=6.5cm]{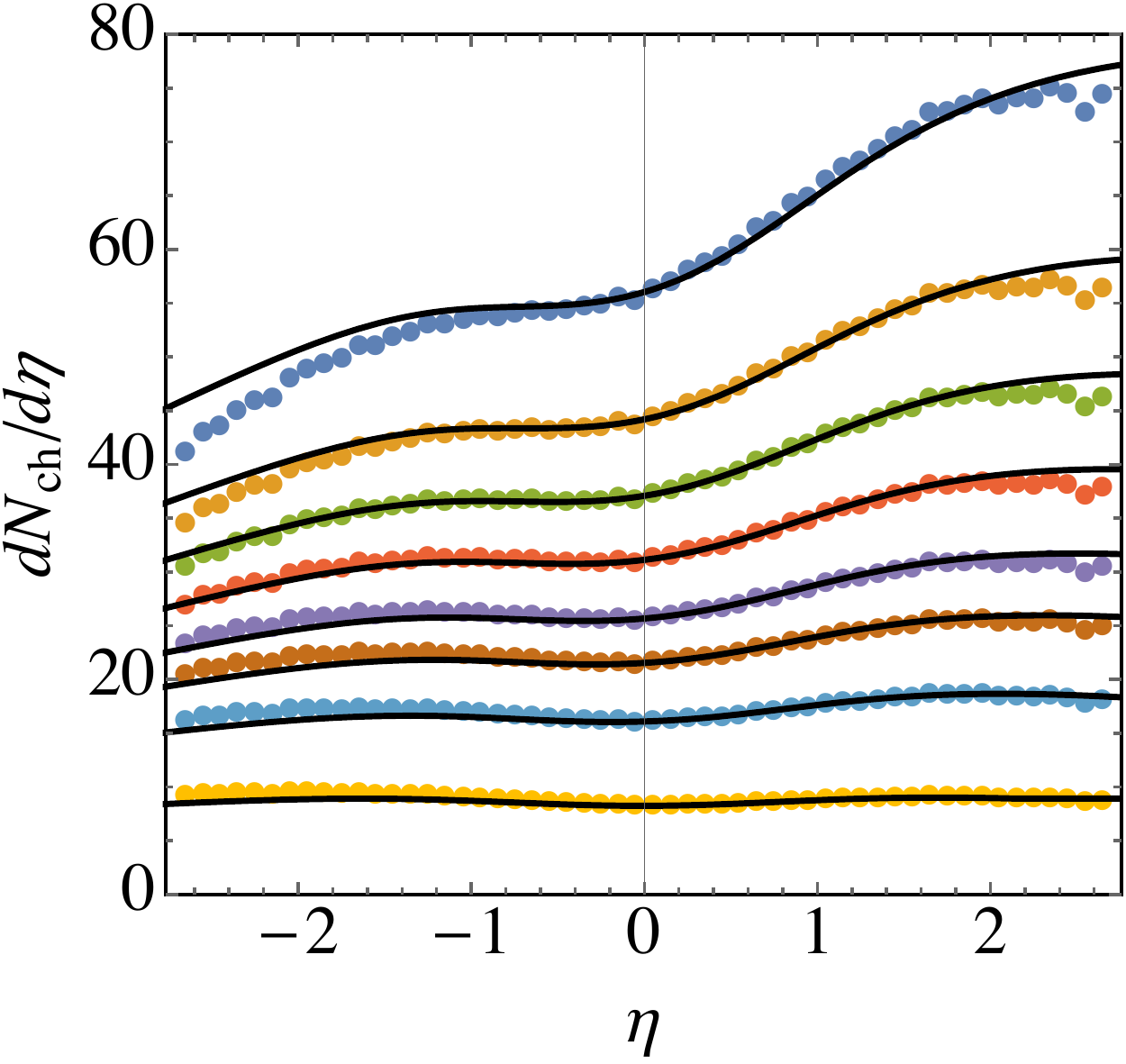}~~~
\includegraphics[width=6.5cm]{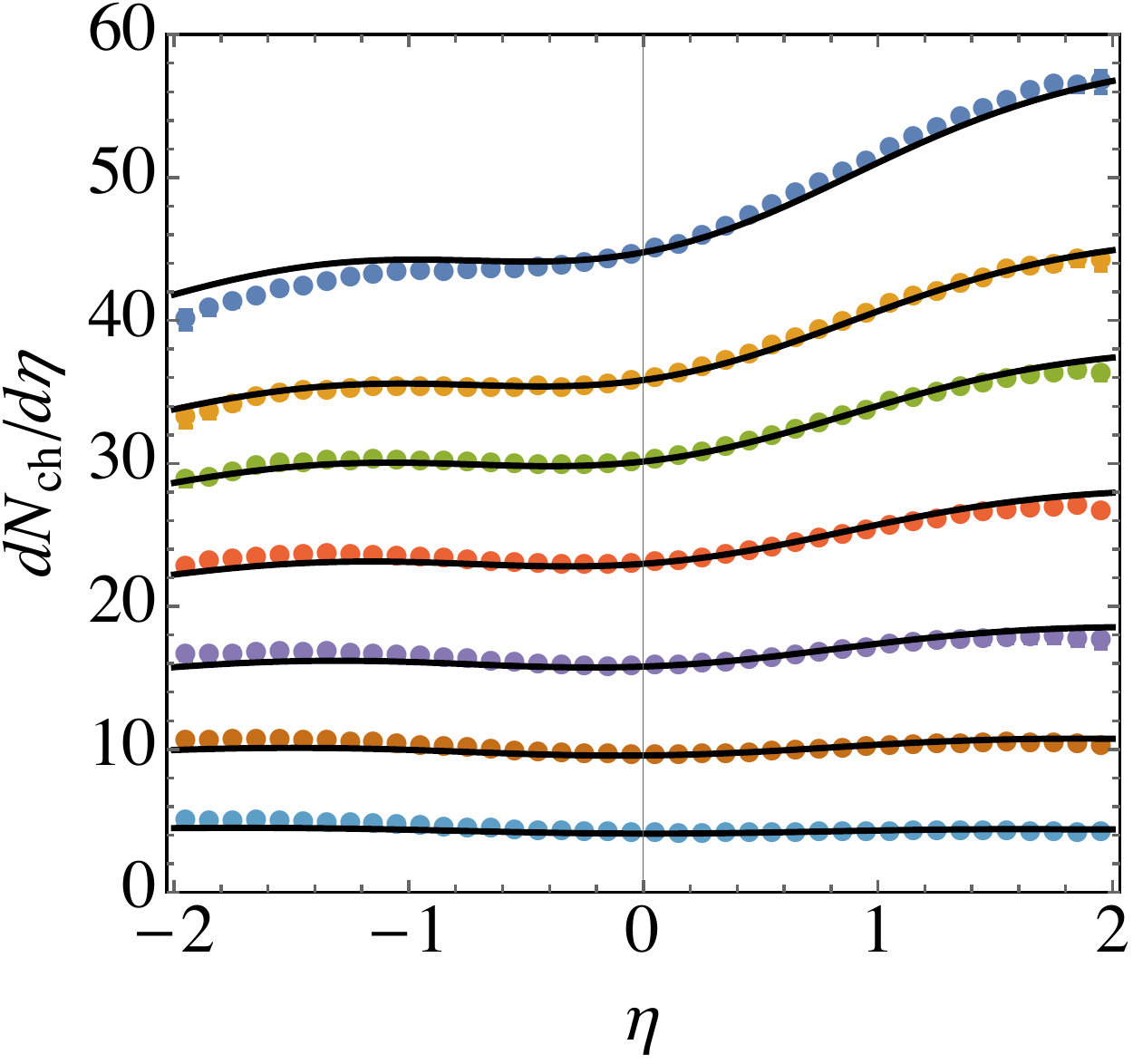}\caption{Multiplicity
dependence on pseudo-rapidity $\eta$ for the fluctuating case with
$\sigma=1.55$. Left plot corresponds to ATLAS whereas the right one to ALICE.
Different curves correspond to the centrality classes defined in
Tables~\ref{tab:ATLAS} and \ref{tab:ALICE}. }%
\label{dNdydatafluct}%
\end{figure}

\begin{figure}[h]
\centering
\includegraphics[width=6.5cm]{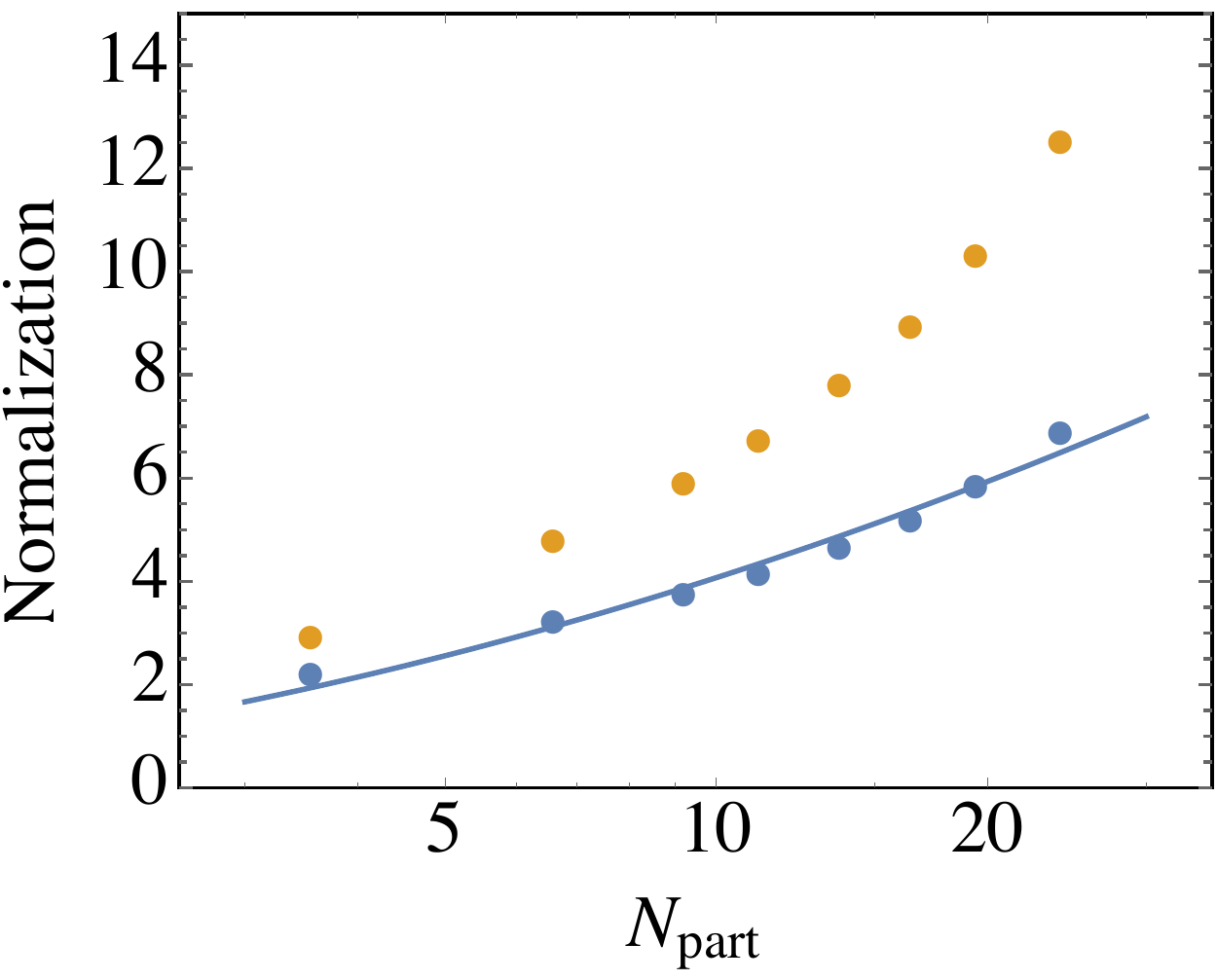}~~~
\includegraphics[width=6.5cm]{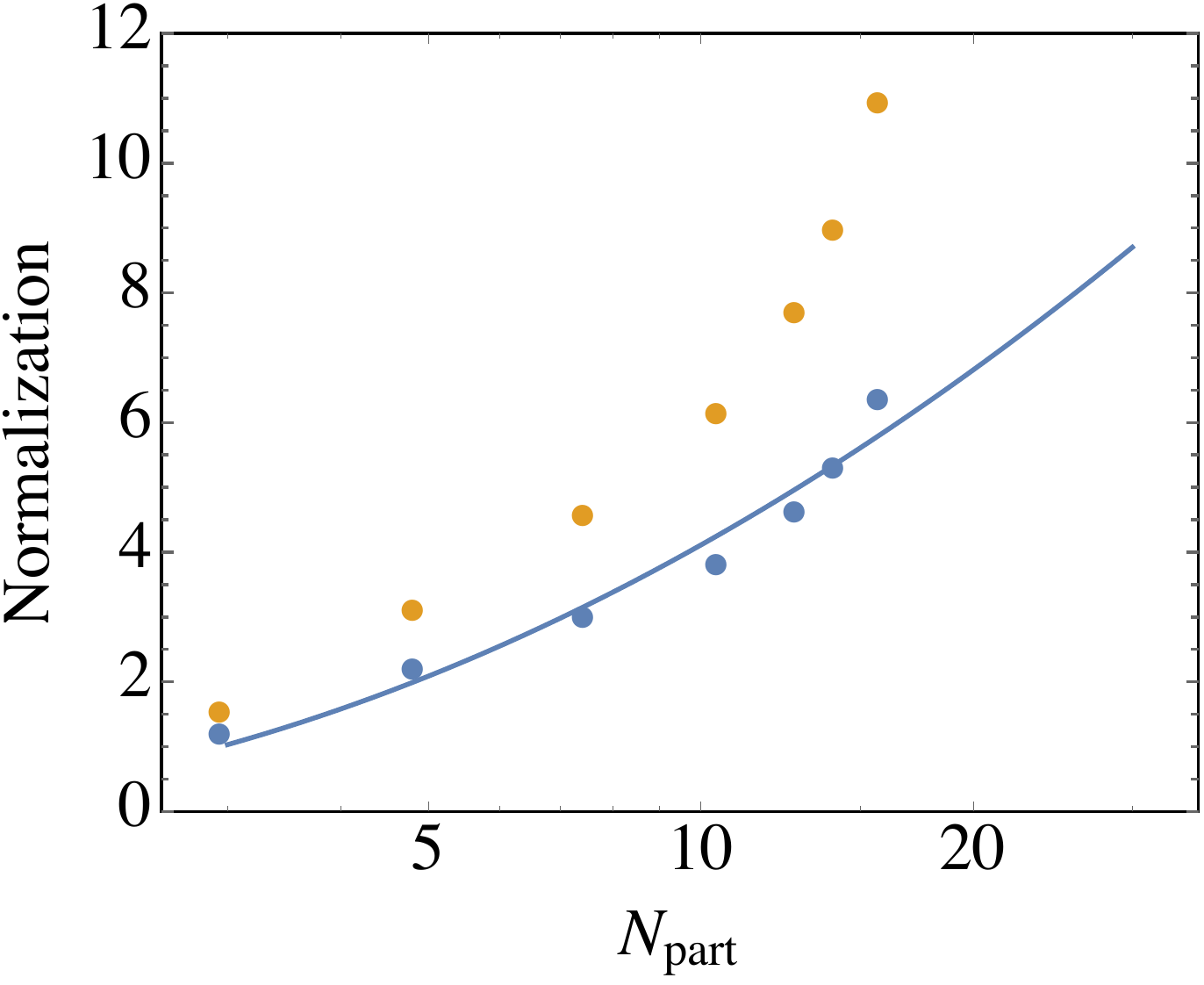}~~~\caption{Normalization of
multiplicity distributions shown in Fig~\ref{dNdydatath} as functions of
$N_{\mathrm{part}}$. Left panel is for ATLAS and right panel for ALICE. Upper
(orange) points correspond to the non-fluctuating case, whereas lower (blue)
points correspond to fluctuations with $\sigma=1.55$. Parametrization of
Eq.~(\ref{STfit}) is shown as a solid (blue) line. }%
\label{NormNpart}%
\end{figure}

Recently ATLAS and ALICE collaborations at the LHC have published multiplicity
distributions in pPb collisions at 5.02 TeV for different centrality classes
summarized in Tables~\ref{tab:ATLAS} and \ref{tab:ALICE}. Throughout this
paper for ATLAS we have used $N_{\mathrm{part}}$ determination by means of the
fluctuating Glauber-Gribov model with $\Omega=0.55$ and for ALICE we use
centrality classes determined by so called V0A method (for details see
Refs.~\cite{atlas,Adam:2014qja}). This data is shown in Figs.~\ref{dNdydatath}
and \ref{dNdydatafluct}. In Fig.~\ref{dNdydatath} we also plot theoretical
predictions for multiplicity distributions without fluctuations corresponding
to Eq.~(\ref{mult1}). We see that CGC predictions are too steep. Moreover,
they are not properly normalized, so we have adjusted the overall factor
$S_{\bot}$ by normalizing the distributions at $\eta=0$. In this way $S_{\bot
}$ does -- as remarked in the Introduction -- depend on $N_{\mathrm{part}}$,
and this dependence is plotted in Fig.~\ref{NormNpart}.

\begin{table}[ptb]
\centering
\begin{tabular}
[c]{|c|c|c|}\hline
class & \% & $\left\langle N_{\text{part}}\right\rangle $\\\hline
c1 & 0-1 & 24.13\\
c2 & 1-5 & 19.42\\
c3 & 5-10 & 16.45\\
c4 & 10-20 & 13.70\\
c5 & 20-30 & 11.16\\
c6 & 30-40 & 9.20\\
c7 & 40-60 & 6.59\\
c8 & 60-90 & 3.55\\\hline
\end{tabular}
\caption{Mean number of participants in different centrality classes as
determined by ATLAS by Glauber-Gribov model with $\Omega=0.55$ }%
\label{tab:ATLAS}%
\end{table}

\begin{table}[ptb]
\centering
\begin{tabular}
[c]{|c|c|c|}\hline
class & \% & $\left\langle N_{\text{part}}\right\rangle $\\\hline
c1 & 0-5 & 15.7\\
c2 & 5-10 & 14.0\\
c3 & 10-20 & 12.7\\
c4 & 20-40 & 10.4\\
c5 & 40-60 & 7.42\\
c6 & 60-80 & 4.81\\
c7 & 80-100 & 2.94\\\hline
\end{tabular}
\caption{Mean number of participants in different centrality classes as
determined by ALICE by so called V0A method.}%
\label{tab:ALICE}%
\end{table}

Next in Fig.~\ref{dNdydatafluct} we superimpose over the data theoretical
predictions of Eqs.~(\ref{i1})--(\ref{sumI}) with fluctuations included.
Theoretical predictions are again normalized at $\eta=0$ and $S_{\bot}$
dependence on $N_{\mathrm{part}}$ is plotted in Fig.~\ref{NormNpart}.

Before we discuss the effects of the fluctuations let us stress that we have
tried to avoid playing with free parameters. Therefore we have kept $\lambda$
fixed to the DIS value of 0.32, we have kept $\eta^{2}_{0}$ fixed to 1.35. The
only free parameter was $\sigma$ and normalization. We have found that the
best description of both ATLAS and ALICE data is for $\sigma=1.55$.

We see that fluctuations make $\eta$ distributions flatter and the
$N_{\mathrm{part}}$ dependence of $S_{\bot}$ is also weaker than in the case
with no fluctuations. The dependence of $S_{\bot}$ can be well approximated by
the following formulae:
\begin{align}
S_{\bot}^{\mathrm{ATLAS}} & =\left(  0.66 + 0.6 \ln(N_{\mathrm{part}})
\right)  ^{2},\nonumber\\
S_{\bot}^{\mathrm{ALICE}} & =\left(  0.09 + 0.84 \ln(N_{\mathrm{part}})
\right)  ^{2}\label{STfit}%
\end{align}
that take into account cross-section fluctuations with $N_{\mathrm{part}}$
discussed in the Introduction. The strength of these fluctuations is, however,
very different in the case of ATLAS and ALICE.

\section{Conclusions}

The original idea of studying multiplicity fluctuations in p$A$ collisions was
that they would allow probes of varying density regions inside the nucleus.
The results we present in this paper suggest that the gluon content of the
proton also itself changes when the multiplicity of final state particles
varies. The analysis we present here attempts a semi-quantitative comparison
to experimental data. This data is not simply explained by the simple
applications of the theory of the Color Glass Condensate or of the Wounded
Nucleon model.  In both cases the shape of
charge particle rapidity distribution is determined by the participant number
$N_{\mathrm{part}}$, and in both cases the simplest models of such a
dependence fail to describe the data. 

The idea that cross sections might fluctuate as an explanation for various
phenomena in heavy ion collisions is not new. This was used in the
Glauber-Gribov implementation to compute the number of participants in the
ATLAS results described above. Motivation was found for the use of
hydrodynamics in high multiplicity p$A$ events at the LHC
\cite{Coleman-Smith:2014daa}. There have also been suggestions that this might
be required from attempts to describe the ridge in p$A$ collisions
\cite{Dusling:2013oia}. Furthermore fluctuations have been implemented
phenomenologically in a Monte Carlo study of p$A$ collisions done in
Ref.~\cite{Dumitru:2011wq} where the authors consider multiplicity variations
for fixed $N_{\mathrm{part}}$ given by the negative binomial distribution
(NBD) that is obtained in the glasma model \cite{Gelis:2009wh,Tribedy:2010ab}.
There our formula (\ref{mult1}) corresponds to the average number of particles
entering NBD. It is, however, difficult to assess the energy dependence of
these fluctuations as it stems from the energy dependence of the NBD itself
and of the impact parameter profile that selects given $N_{\mathrm{part}}$
(that in a sense would correspond to the fluctuations of the nucleus
saturation scale).

On the other hand, the arguments for a fluctuating saturation momentum have
been primarily theoretical \cite{Munier:2003sj}%
--\nocite{Munier:2003vc,Iancu:2004es,Hatta:2006hs}\cite{Dumitru:2007ew}.
Interest in such fluctuations diminished somewhat when no evidence of
diffusive scaling has been found in DIS
\cite{Gelis:2006bs,Praszalowicz:2013iyi} and after it was argued that the
evolution of the strength of such fluctuations was very weak with varying
rapidity. 
 However, if the fluctuations are largely rapidity independent -- as
it is the case of the present paper -- geometrical scaling variable $\tau$ is
not modified and there is no contradiction between our results and the absence
of diffusive scaling in DIS. Another point is the energy dependence of such
fluctuations. This needs to be tested in comparative studies at both RHIC and
LHC energies.

Our finding that the fluctuations can account for the data must be taken with
much caution. It is simply a first attempt and the number extracted for the
width of the fluctuations of the logarithm is 
rather large, and exceeds by a factor of 1.5 the one of recent
analysis of Ref.~\cite{tribedy}.  This may be
because in our simple analysis, we do not consider fluctuations of the
saturation momentum of the nucleus. It is not obvious how to do this, since
the fluctuations would be local over some transverse scale size of order the
inverse saturation momentum. Similarly, we expect there to be local structure
in the proton since the size scale of the saturation momentum is small
compared to the proton size. The issue of a local fluctuating saturation
momentum was addressed by Iancu and McLerran \cite{Iancu:2007st} within the
context of conformal field theory, but there is as yet no practical
implementation of such ideas useful for understanding experimental data. 
Furthermore, in our simple analysis we have
neglected other sources of fluctuations that have been taken into account in
Ref.~\cite{tribedy}.   Finally, since we do not include
any dumping for large $y$'s (as has been done {\em e.g.} in Ref.~\cite{Kharzeev:2004if}
where large $x$ suppression for gluon densities has been included),
our calculation is most reliable in the mid rapidity region.
Evidently a more realistic model must contain such
effects. 

If we accept the idea of a fluctuating saturation momenta as a working
hypothesis, there are a variety of tests one can imagine. Fluctuations in the
multiplicity for very high multiplicity predicted in the IP-glasma model will
be modified by such effects \cite{Schenke:2012hg,tribedy}. The correlation
between multiplicity and jet production will also be changed, since higher
saturation momenta in the proton means more gluons and therefore greater
probability of producing a jet.

The original idea of studying multiplicity fluctuations in p$A$ collisions was
that they would allow probes of higher density regions inside the nucleus. The
results we present in this paper suggest that the gluon content of the proton
also itself changes when the multiplicity of final state particles varies.

\section*{Acknowledgements}

We would like to thank A. Bialas, A. Bzdak, E. Iancu, D. Kharzeev, E. Levin,
P. Tribedy and R. Venugopalan for discussion and remarks. The authors are
supported under Department of Energy contract number Contract No.
DE-SC0012704. Research of MP has been supported by the Polish NCN grant 2014/13/B/ST2/02486.

\end{document}